\documentclass{JCIN22}
\usepackage[super]{gbt7714}
\usepackage{wrapfig}
\usepackage{etoolbox}
\makeatletter
\patchcmd{\@maketitle}
  {\begin{tabular}[t]{@{}l@{}}}   
  {\begin{tabular}[t]{@{}c@{}}}   
  {}{}
\makeatother

\begin{document}

\ArticleType{Review paper}
\Year{2025}

\title{Meta-Backscatter: Long-Distance Battery-Free Metamaterial-Backscatter Sensing and Communication}

\author[ ]{Taorui~Liu}
\author[ ]{Xu~Liu}
\author[ ]{Zhiquan~Xu}
\author[ ]{Houfeng~Chen}
\author[ ]{Hongliang~Zhang}
\author[ ]{Lingyang~Song}

\maketitle

{\bf\textit{Abstract---}
Battery-free Internet of Things (BF-IoT) enabled by backscatter communication is a rapidly evolving technology offering 
advantages of low cost, ultra-low power consumption, and robustness.
However, the practical deployment of BF-IoT is significantly constrained by the limited communication range of common 
backscatter tags, which typically operate with a range of merely a few meters due to inherent round-trip path loss.
Meta-backscatter systems that utilize metamaterial tags present a promising solution, retaining the inherent advantages 
of BF-IoT while breaking the critical communication range barrier. 
By leveraging densely paved sub-wavelength units to concentrate the reflected signal power, metamaterial tags enable a 
significant communication range extension over existing BF-IoT tags that employ omni-directional antennas.
In this paper, we synthesize the principles and paradigms of metamaterial sensing to establish a unified design 
framework and a forward-looking research roadmap.
Specifically, we first provide an overview of backscatter communication, encompassing its development history, working 
principles, and tag classification. 
We then introduce the design methodology for both metamaterial tags and their compatible transceivers. 
Moreover, we present the implementation of a meta-backscatter system prototype and report the experimental results based 
on it. 
Finally, we conclude by highlighting key challenges and outlining potential avenues for future research.
\\[-1.5mm]

\textit{Keywords---}
meta-backscatter, metamaterial tags, battery-free Internet of Things, communication distance}

\section{Introduction}
\label{Introduction}
The Internet of Things (IoT) will serve as the core enabling technology for sixth-generation (6G) communications, 
supporting critical applications in healthcare, intelligent industry, and environmental protection \cite{52, 53}. 
With projected tag deployments approximately ten times greater than current 5G networks \cite{54}, 6G IoT systems 
demand tags that are low-cost, ultra-low power, and robust, i.e., capable of working in harsh environments for a long 
time without maintenance \cite{c2}. 
However, conventional tags fail to meet these requirements due to their reliance on expensive and delicate electronic 
components such as transceivers, modulators, and processing units, combined with power-intensive battery systems that 
become unsustainable at scale \cite{55}. 
Therefore, realizing 6G IoT networks necessitates innovative tag architectures that fundamentally address these 
convergent challenges.

To address these limitations of traditional IoT tags, battery-free IoT (BF-IoT) tags have emerged, employing backscatter 
communication technology to eliminate internal power requirements \cite{1}. 
Rather than generating active transmissions, these devices modulate and reflect incident radio frequency signals toward 
transceivers \cite{57}. 
These devices encode data through controlled manipulation of load impedances, which alter scattering coefficients to 
modulate reflected signal amplitude and phase \cite{58}. 
This approach offers significant advantages: ultra-low power consumption at $\mu$W levels, simplified architecture 
consisting only of antennas and radio control modules, manufacturing costs reduced to cents per unit, and operational 
lifespans exceeding ten years \cite{59}. 
The technology is particularly valuable for deployment in harsh environments where maintenance is impractical, including 
mining operations, chemical processing facilities, and embedded building infrastructure applications, making backscatter 
tags an optimal solution for long-term, large-scale sensing networks in challenging conditions \cite{56}. 
This innovative approach thus represents a significant advancement in overcoming the fundamental constraints of 
traditional IoT tags.

Despite the numerous advantages of backscatter tags, due to the inherent round-trip path loss of backscatter 
communication, the communication range of backscatter tags is significantly shorter compared to traditional IoT devices 
\cite{47, 48}. 
The typical communication range for common backscatter tags is only up to a few meters \cite{c5, 45, 46}, while the 
Third Generation Partnership Project (3GPP) requirements specify that IoT devices should achieve communication ranges 
exceeding 10-meter level.
While some backscatter tags employing specially designed excitation signals, such as LoRa backscatter based on Chirp 
Spread Spectrum (CSS), can achieve a longer communication range, their power consumption is not low enough to support 
battery-free applications \cite{48}.
This limited communication range significantly restricts the practical deployment of backscatter tags.
To address the limitations of existing backscatter tags, metamaterial tags have been proposed. 
Specifically, metamaterial tags are sub-wavelength resonators combining environmentally sensitive structure that are 
periodically arranged on dielectric substrates \cite{b2, 42}. 
Unlike existing backscatter tags that typically employ omni-directional antennas with low directional gain \cite{b6}, 
the densely paved units with sub-wavelength spacing result in more concentrated reflected signal power and thus longer 
communication ranges, according to antenna theory \cite{b3, 71}. 
By combining the low cost, low power consumption, and high robustness of backscatter tags while overcoming communication 
range limitations, metamaterial tags demonstrate great potential for widespread applications \cite{10, 43}. 

However, designing efficient sensing systems based on metamaterial tags, refered to as \textit{meta-backscatter 
systems}, presents significant challenges due to the simultaneous sensing and transmission requirements. 
The frequency response characteristics of the tag critically influence both its sensing accuracy and communication 
performance \cite{12}, and transmission effects cannot be disregarded during the sensing process \cite{24}. 
These aspects have been largely overlooked in existing reviews on backscatter communication \cite{47, 57, 40}.
Moreover, current reviews on metamaterial tags \cite{11, 75, 76, 77, 43} have primarily focused on the design of 
sensing functionalities, often neglecting essential considerations related to the transmission process. 
Furthermore, they typically catalogue and summarize individual designs without proposing a comprehensive system 
framework that integrates tag design with custom transceiver implementation. 
Consequently, establishing a unified framework for the co-design of sensing and communication is imperative.

Specifically, this work addresses the above gap by summarizing the fundamental principles, design paradigms, and key 
techniques specific to metamaterial sensing. 
Our main contributions are as follows:
\begin{itemize}
\item First, we present the fundamental principles of metamaterial tags and establish an equivalent circuit model to 
facilitate their design. 
A comprehensive review of existing tag designs further complements this theoretical groundwork. 
\item Next, we establish a signal sensing and transmission model for meta-backscatter systems. 
Based on this model, we analyze key signal detection techniques, highlighting their applicability and specific 
requirements.
\item Furthermore, we implement a meta-backscatter system prototype and provide corresponding experimental results to 
demonstrate its practical feasibility.
\item Finally, we outline the major research challenges and future directions, aiming to identify open problems and 
stimulate further investigation in this field.
\end{itemize}

The remainder of this paper is organized as follows. 
Section~\ref{Overview of Backscatter} provides an overview of backscatter communication, covering development history, 
fundamental principles, as well as tag category comparisons.
Section~\ref{Metamaterial Tag Design} examines design principles, SRR-based circuit modeling, and a comprehensive 
literature review of metamaterial tags.
Section~\ref{Signal Transmission and Detection} introduces a signal sensing and transmission model, followed by a 
discussion of key signal detection techniques.
Section~\ref{Implementation and Prototype} presents the implementation of a meta-backscatter system prototype and 
corresponding experimental results. 
Section~\ref{Challenges and Open Questions} identifies existing challenges and future research directions, while 
Section~\ref{Conclusion} concludes the paper. 

\section{Overview of Backscatter} 
\label{Overview of Backscatter}
This section begins with introduction of the development history of backscatter communication.
Then we illustrate the fundamental principles of backscatter systems.
Subsequently, we categorize backscatter tags by their signal modulation schemes and discuss their respective advantages 
and disadvantages.

\subsection{Development History of Backscatter}
The origins of backscatter communications can be traced to early radar systems used to detect the location of the enemy 
aircraft during World War II \cite{b1}. 
In 1948, Harry Stockman published what is widely recognized as the first formal conceptualization of backscatter 
communications in his landmark paper, ``Communication by Means of Reflected Power'' \cite{1}, thereby establishing 
the theoretical foundation for subsequent developments in this field. 
Following this seminal work, backscatter communications continued to evolve along multiple research trajectories, 
ultimately culminating in the successful development of radio frequency identification (RFID) technology \cite{2}. 
The 1970s witnessed significant milestones, including the presentation of the earliest RFID prototype in patent form 
\cite{o1} and a notable breakthrough by Los Alamos Scientific Laboratory, which developed an electronic 
identification system utilizing modulated backscatter waves from an ID tag \cite{3}. 
These developments marked the transition of backscatter communications from theoretical concept to practical reality.

As RFID technology matured, the first generation of backscatter tags emerged to eliminate battery dependence in 
wireless tags \cite{c1, 4}. 
In this typical configuration, conventional tags are integrated into RFID tags, enabling the backscattered signals to 
carry sensing data. 
Since these systems operate without batteries, both the RFID chip and tag rely on energy harvesting for power. 
However, in first-generation backscatter tags, the sensing and communication modules function independently. 
To fully exploit the inherent properties of backscatter communications, researchers developed second-generation 
electromagnetic RFID tags. 
Unlike their predecessors, these tags incorporate sensitive materials directly into the tag's antenna rather than 
using conventional sensing components \cite{5}. 
The sensing data is directly encoded into the RFID tag's electrical parameters, such as input impedance or antenna gain. 
This elimination of power-consuming tags significantly extends the effective operational range.

However, manufacturing and maintenance costs associated with chips remained a significant obstacle to widespread 
backscatter tag adoption. 
In the 2010s, third-generation chipless RFID tags emerged as a research focus due to their potential for reduced 
costs \cite{6, 7}. 
These chipless RFID tags consist essentially of passive antennas or resonators that exhibit resonance frequencies 
sensitive to specific environmental targets, enabling sensing data to be extracted from the spectral characteristics of 
backscattered signals \cite{8, 9}. 
This approach eliminates the need for digital circuits, relying solely on analog signal processing. 

While offering cost advantages, chipless RFID tags suffer from limited accuracy and operational range, constraining 
their practical deployment. 
To address these challenges, a novel chipless technique utilizing metamaterial tags has been proposed \cite{10}. 
These tags feature specialized resonant elements arranged in periodic two-dimensional arrays to achieve enhanced 
sensitivity and improved backscatter beam gain. 
Consequently, metamaterial tags enable long-range sensing without power consumption, representing a promising 
solution for passive Internet of Things applications.

\subsection{Fundamentals of Backscatter}
A backscatter communication system consists of two primary components: transceivers and tags.
The transceiver transmits excitation signals within a specific frequency band and receives scattered signals from tags.
The tag modulates its scattering coefficient based on the sensing information to be transmitted, thereby altering the 
scattered signal.
The operation of a backscatter communication system involves three sequential steps: forward transmission, tag 
scattering, and backward reception.
First, transmitters such as WiFi hotspot or cellular base stations emit wireless excitation signals.
Second, the tag adjusts its scattering coefficient according to the sensing information, thereby modifying the scattered 
signal.
Finally, the receiver estimates the tag's scattering coefficient from the received scattered signal to infer the tag's 
sensing information.

Specifically, scattered signals $S_{\text{out}}$ encode sensing information through variations in amplitude or phase: 
\begin{equation}
    \begin{aligned}
        S_{\text{out}} = \left| S_{\text{in}} \right| \left| \Gamma_{\text{tag}} \right| \mathrm{e}^{\mathrm{i}(2\pi 
        f_{\text{in}}t + \theta_{\text{in}} + \theta_{\text{tag}})} \text{,} 
    \end{aligned}
\end{equation}
where $\left| S_{\text{in}} \right|$ and $\theta_{\text{in}}$ denote the amplitude and phase of the excitation signal, 
$f_\text{{in}}$ denotes the frequency of the excitation signal, and $\left| \Gamma_{\text{tag}} \right|$ and 
$\theta_{\text{tag}}$ denotes the amplitude and phase of the tag's scattering coefficient. 
Based on whether the scattering coefficient exhibits discrete or continuous variations, tags can be classified as 
digital or analog types, which will be discussed in the following.

\subsection{Classification of Backscatter Tags}
\begin{figure}[!t]
    \centering
    \begin{minipage}[c]{\textwidth}
    \includegraphics[width=3.5in]{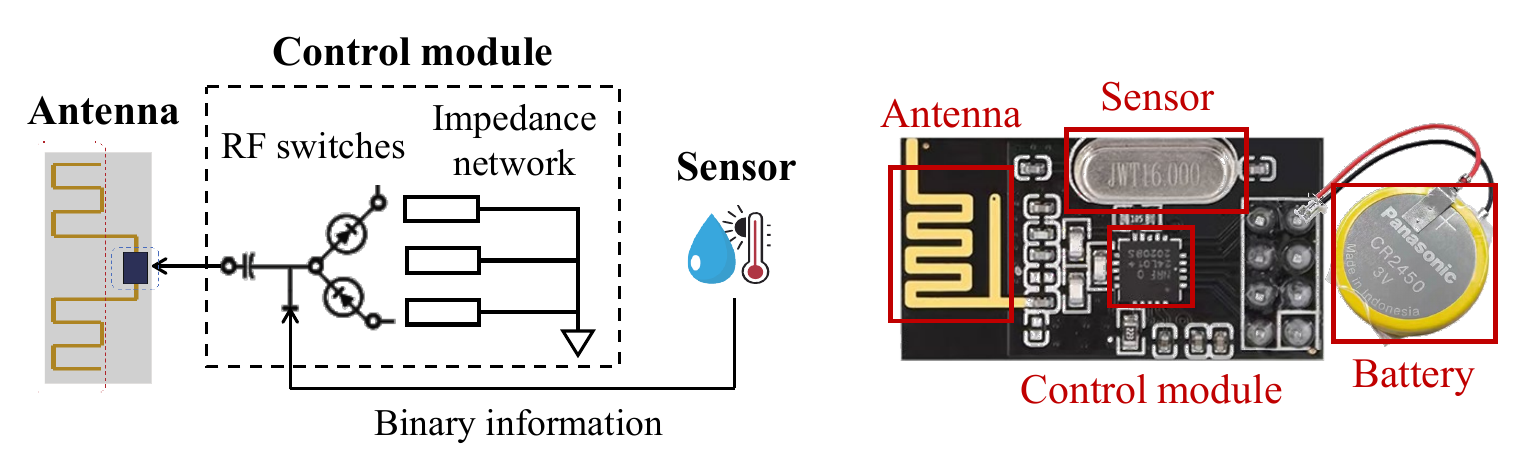}
    \end{minipage}
    \begin{minipage}[c]{\textwidth}
    \includegraphics[width=3.5in]{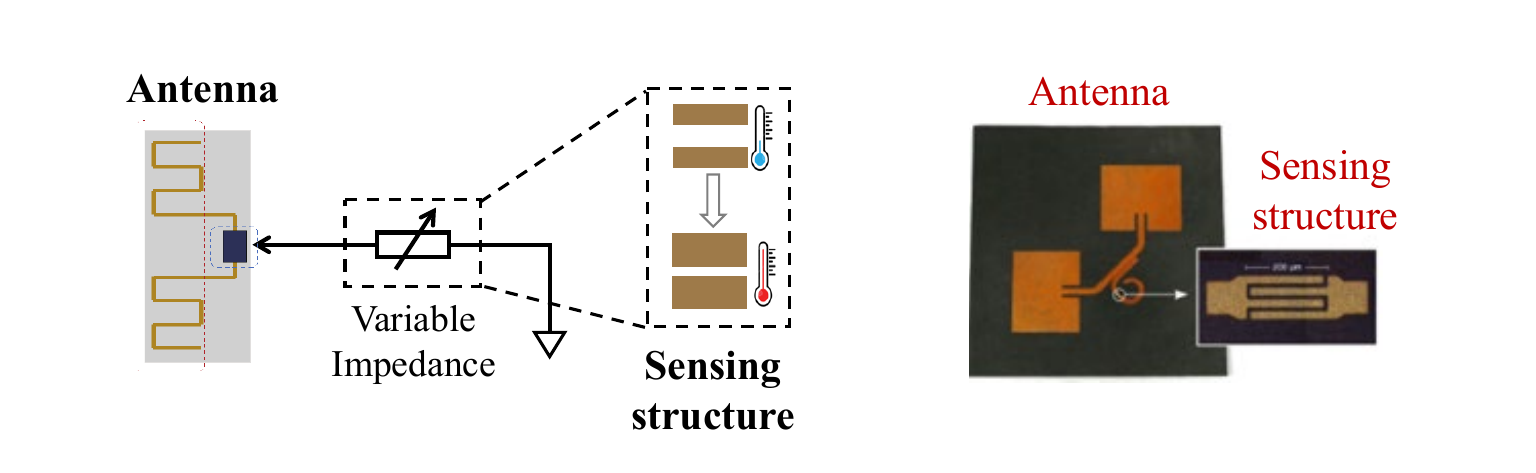}
    \end{minipage}
    \caption{(a) A digital backscatter tag. (b) An analog backscatter tag.}
    \label{classification}
\end{figure}

\subsubsection{Digital Backscatter Tags}
As illustrated in Fig. \ref{classification}(a), a digital backscatter tag comprises four key components: a peripheral 
tag, a signal control module, a radio frequency (RF) antenna, and a power supply (either battery or energy harvesting 
module). 
The peripheral tag detects environmental information and converts them into a binary bit stream, which is then 
transmitted to the signal control module. 
This module consists of RF switches and an impedance network, where the binary bit stream from the tag controls the 
switching states of the RF switches, effectively connecting the RF antenna to different load impedances. 
The impedance matching or mismatching between the antenna and load alters the tag's scattering coefficient, enabling 
data modulation onto the incident RF signal. 
Through appropriate demodulation techniques, the transceiver can successfully recover the transmitted sensing data. 
The power supply module provides the necessary energy for all operational processes.
The switchable impedance values of the network are limited, consequently, the variation of the scattering coefficients 
is also discrete.

Various commercially available chip-based RFID tags fall under the category of digital backscatter tags. 
Due to the superior interference resistance of digital modulated signals, digital tags can achieve higher sensing 
accuracy \cite{b7}.
However, energy harvesting modules typically provide power only in the microwatt range \cite{44}, while size-constrained 
batteries employ ultra-thin designs that cannot sustain high power consumption over extended periods. 
Consequently, digital tags can accommodate only a limited variety of tags and support basic sensing functions such as 
temperature or light intensity monitoring \cite{c1,4}. 
Furthermore, the delicate microchips in digital tags may incur prohibitive maintenance costs in harsh environmental 
conditions or during large-scale deployments \cite{40, 41}.

\subsubsection{Analog Backscatter Tags}
Unlike digital tags, analog tags consist solely of an RF antenna and an electromagnetic resonant structure, as shown in 
Fig. \ref{classification}(b). 
During measurement, the transceiver transmits a frequency sweep signal. 
The electromagnetic resonant structure resonates with the incident signal, absorbing signals at specific frequency 
points. 
As the environmental parameters being sensed change, the impedance of the electromagnetic resonator structure varies 
accordingly, causing shifts in the absorption frequency points. 
The transceiver then extracts sensing information from the scattered signal spectrum. 
Since the impedance of the resonant structure changes continuously with environmental variations, the scattering 
coefficient also varies continuously.

Chipless RFID tags are among the most widely studied analog tags. 
The sensing capability of analog tags is achieved through electromagnetic resonant structures without requiring batteries 
or energy harvesting, thereby enabling support for a wider range of sensing functions, including strain monitoring, 
humidity monitoring, gas concentration detection, or gesture recognition \cite{21, 20, 38, 39}. 
Furthermore, analog tags feature simple structures without delicate microchips, resulting in lower production and 
maintenance costs \cite{35}. 
However, analog-modulated signals are susceptible to channel noise interference, resulting in limited sensing accuracy
\cite{b7}.

The replacement of the omni-directional antenna in existing backscatter tags with a metamaterial array enables a 
substantial extension of communication range and, crucially, provides support for both digital and analog modulations. 
Owing to this unique design philosophy, the discussion in subsequent sections will focus on issues pertaining to analog 
metamaterial tags.

\section{Metamaterial Tag Design}
\label{Metamaterial Tag Design}
This section begins by elucidating the design principles of metamaterial tags. 
Subsequently, we present the equivalent circuit model, using split-ring resonator (SRR) as an example, which is one of 
the most prevalent metamaterial structures. 
The section concludes with an overview of metamaterial designs from related works.

\subsection{Design Principles}
\begin{figure}[!t]
    \centering
    \includegraphics[width=3.0in]{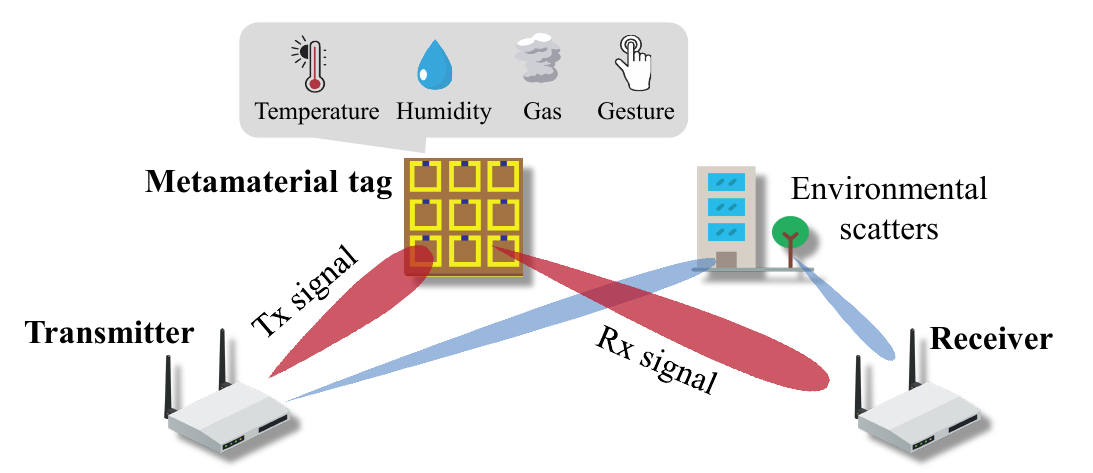}
    \caption{The illustration of a meta-backscatter system.}
    \label{system}
\end{figure}
Fig. \ref{system} illustrates the composition of a meta-backscatter system.
As discussed in the previous section, metamaterial tags enable simultaneous sensing and transmission through 
frequency-selective reflection of signals based on their structural design. 
This operational principle necessitates an approach to design metamaterial tags, requiring joint consideration of 
sensing and transmission performance \cite{12}. 
Specifically, the sensing accuracy is intrinsically linked to variations in reflected signals across different tag 
states, while the tag structure significantly influences the intensity of received signals, thereby affecting the 
overall transmission range. 
Furthermore, the metamaterial structure incorporates multiple parameters, influencing the tag performance jointly 
\cite{24}. 
Consequently, the optimal tag design necessitates a comprehensive evaluation of all relevant structural parameters to 
identify the most effective configuration within a vast domain of feasible options.

To address the aforementioned challenges, we propose a multi-objective optimization framework for metamaterial 
structures. 
This framework aims to enhance two critical aspects of system performance: sensing accuracy and transmission range. 
By striking a balance between these objectives, the framework seeks to achieve optimal overall functionality. 
Sensing accuracy, primarily dependent on the tag's sensitivity, is quantified through measurable changes in the reflected 
signal. 
The sensing principle of metamaterial relies on their resonant frequency's sensitivity to specific targets due to 
the presence of sensitive materials \cite{b2}, and information about sensing targets can be obtained by measuring the 
resonant frequency of reflected signals \cite{21}. 
Consequently, a larger frequency shift generally indicates higher sensing accuracy \cite{20}. 
The Q-factor, or quality factor, is a dimensionless parameter characterizing the efficiency of a resonant system 
\cite{b5}. 
A higher Q-factor signifies a lower rate of energy loss relative to stored energy, implying more efficient and sharper 
resonance. 
At equal resonance frequency shifts, a higher Q-factor results in more distinguishable changes in the reflected signal, 
thereby enhancing sensing accuracy \cite{19}. 
For these reasons, this study adopts resonance frequency shift and Q-factor as metrics for sensing accuracy.

Concurrently, the transmission range is evaluated using the signal-to-noise ratio (SNR), which requires careful 
consideration of the reflected signal's intensity. 
While some researchers have used change gradient in reflected power \cite{12,18} or average reflected power \cite{10} to 
measure transmission range, we employ the minimum reflected power over the frequency band. 
A higher minimum reflected power corresponds to a higher minimum SNR over the frequency band, thus enabling a longer 
transmission range \cite{b3}. 
Besides, to simplify the multi-parameter optimization, we utilize equivalent circuit model, which is discussed in detail 
in the subsequent section. 

\subsection{Equivalent Circuit Model}
The metamaterial structure incorporates numerous design parameters, including unit cell geometry, spacing, and material 
composition, all of which collectively determine the tag's frequency response, as discussed in the previous section.
Attempting to optimize all these parameters simultaneously would result in prohibitive computational complexity, 
rendering the process impractical. 
To address this challenge, a frequency response model based on an equivalent circuit has been proposed, striking a 
balance between computational efficiency and modeling precision \cite{14}. 
This approach simplifies the metamaterial tag structure by representing it as a limited number of circuit components, 
allowing for more manageable analysis. 
By projecting parameter changes onto these circuit components, the model intuitively demonstrates their impact on the 
overall frequency response. 
The optimization process is further streamlined through the classification of parameters and the selection of those 
deemed essential \cite{13}. 
Consequently, this methodology significantly reduces the feasible set for optimization, making the process more 
tractable and efficient while maintaining a high degree of accuracy in modeling the frequency response.

\begin{figure}[!t]
    \centering
    \includegraphics[width=2.5in]{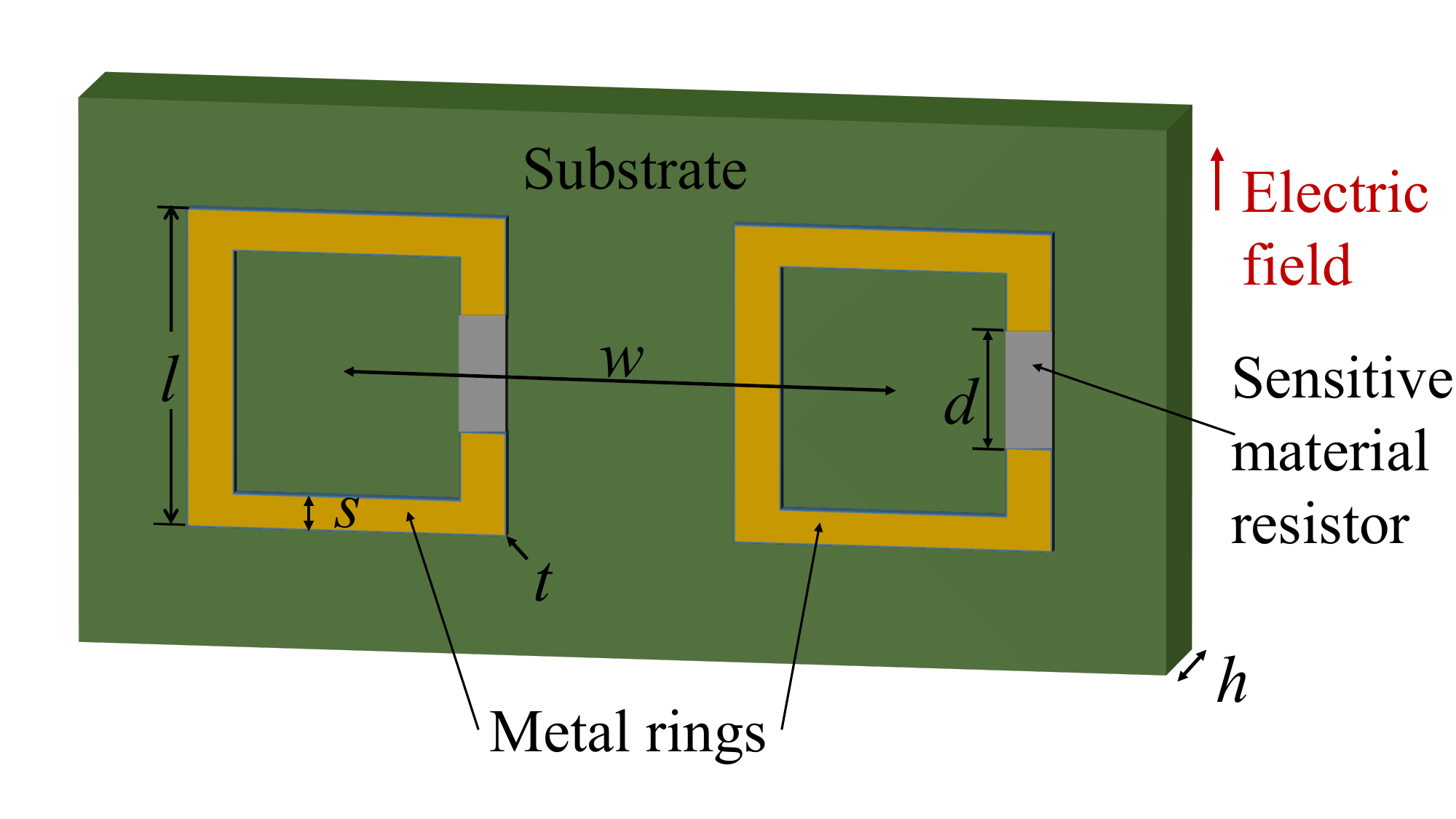}
    \caption{SRR particles.}
    \label{SRR}
\end{figure}

Split-ring resonators (SRRs) are among the most frequently utilized metamaterial tag structures due to their ease of 
design and fabrication \cite{11}. 
As shown in Fig. \ref{SRR}, the structural dimensions of the SRR particle can be described as follows: We use $l$ to 
denote the side length of the ring, $d$ to denote the width of the gap, $s$ to denote the width of the ring, $w$ to 
denote the distance between the centers of adjacent SRR particles, $t$ to represent the thickness of the ring, and $h$ 
to denote the thickness of the substrate. 
It is worth noticing that the direction of the gap is aligned vertically, and the direction of the electric field is 
parallel to the gap. 
Using SRRs as an example, we present the proposed equivalent circuit model and provide insight for metamaterial tag 
design based on this model in the following.

\subsubsection{Overview of model}
\begin{figure}[!t]
    \centering
    \includegraphics[width=3.5in]{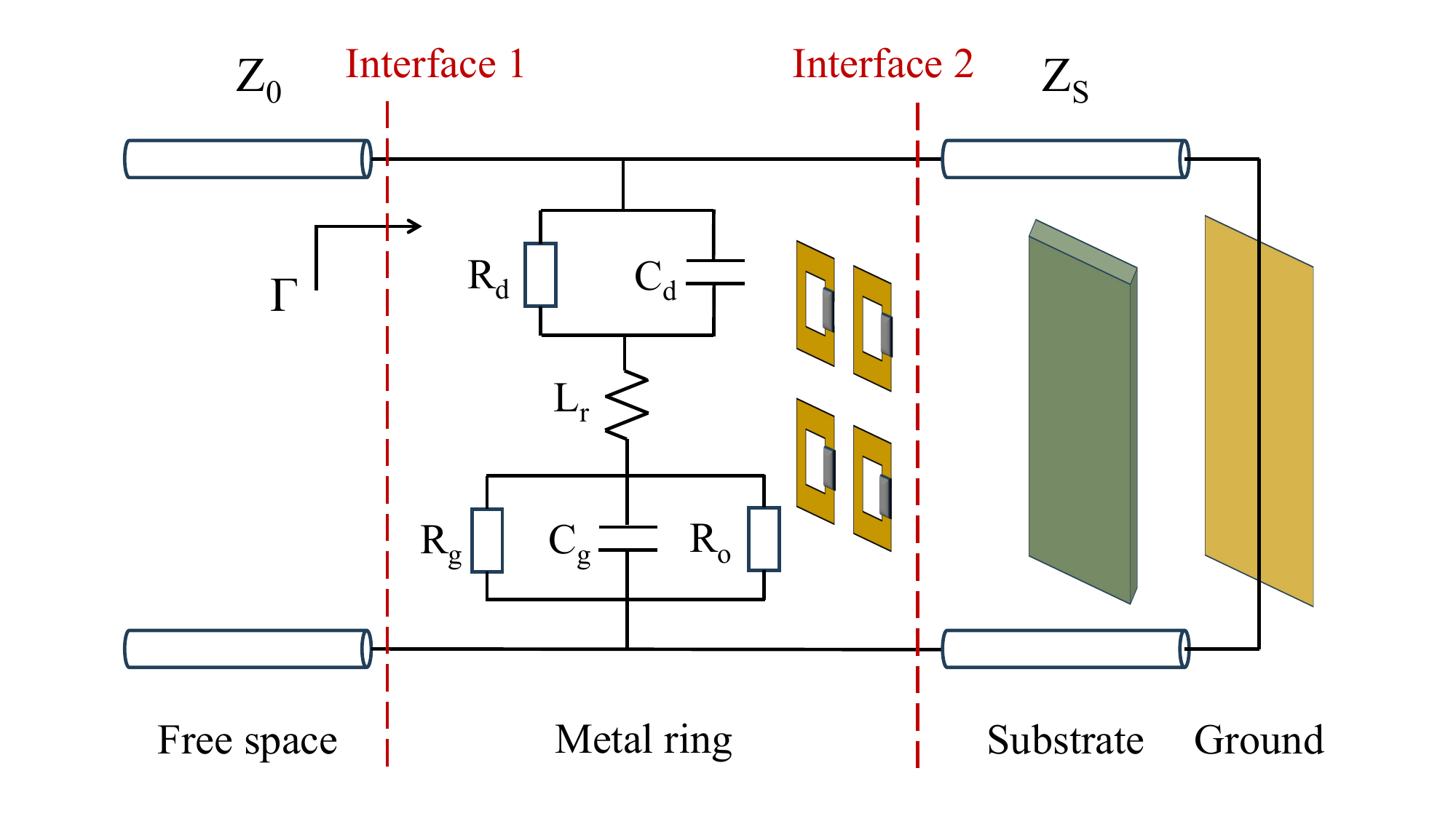}
    \caption{Equivalent circuit model.}
    \label{circuit}
\end{figure}
The layout of the investigated metamaterial tag structure and its corresponding equivalent circuit are illustrated in 
Fig. \ref{circuit}. 
The structure comprises SRRs with environmentally sensitive material resistors straddling their gaps, periodically 
arranged on a lossy grounded dielectric substrate. 
The SRRs on the upper surface are analogous to a two-port circuit network, while the substrate and metal ground on the 
reverse side are equivalent to a terminated short-circuited transmission line. 
According to transmission line theory, the scattering coefficient $\Gamma$ of the metamaterial tag, can be calculated 
as \cite{b4}: 
\begin{equation}
    \begin{aligned}
        \Gamma=\frac{Z_{\text{tag}}(f, \bm{\psi}_{\text{env}}) - Z_0}{Z_{\text{tag}}(f, \bm{\psi}_{\text{env}}) +Z_0} 
        \text{,} \label{scattering coefficient}
    \end{aligned}
\end{equation}
where $Z_0=377\ \Omega$ is equivalent impedance for free-space transmission \cite{b4}, $\bm{\psi}_{\text{env}}$ denotes 
the environmental parameter to be sensed, and $Z_{\text{tag}}$ is equivalent impedance of the metamaterial tag, i.e., 
impedance from interface 1 to the right as shown in Fig. \ref{circuit}. 
Based on derivations in \cite{b4}, we can prove that $Z_{\text{tag}}$ is equal to the parallel connection of 
$Z_{\text{SRR}}$ and $Z_S$, where $Z_{\text{SRR}}$ is the total impedance of the RLC circuit between interfaces 1 and 2, 
$Z_S$ is impedance of the terminated short-circuited transmission line from interface 2 to the right as shown in Fig. 
\ref{circuit}. 

\subsubsection{Circuit components}
Next, we focus on the the RLC circuit between interfaces 1 and 2 as shown in Fig. \ref{circuit}. 
In the presented RLC circuit, $C_d$ is the dielectric capacitor, $R_d$ denotes the dielectric resistor, $L_r$ denotes 
the metal ring inductor, $C_g$ is the gap capacitor, $R_g$ is the gap resistor, and $R_o$ denotes the ohmic resistor. 
These equivalent circuit parameters are described as follows:

$C_d$ represents the capacitance formed between adjacent SRRs. As each SRR is surrounded by four others, $C_d$ can be 
expressed as a series connection of four capacitors \cite{16}, such that $C_d=C_0/4$, where $C_0$ denotes the capacitance 
between two neighboring SRRs and can be modeled as described in \cite{16}. 
Similarly, $C_g$ denotes the capacitance formed at the gap of the metal ring, which can be calculated using the same 
method as $C_0$.

The imaginary component of the substrate's relative dielectric constant introduces a resistive element that accounts for 
the effect of the lossy substrate in proximity to the metal rings. 
This loss component can be represented by a resistor $R_d$ in parallel with the lossless capacitor $C_d$, and a resistor 
$R_g$ in parallel with the lossless capacitor $C_g$ \cite{14}.

$L_r$ represents the inductance formed by the metal ring, which can be modeled based on the method described in \cite{17}. 
The value of $R_o$ corresponds to the resistance of the sensitive material resistors.

\subsubsection{Discussions based on circuit model}
Consider the above RLC circuit in parallel with the substrate impedance $Z_S$. Let $R_{\text{total}}$, 
$L_{\text{total}}$, and $C_\text{{total}}$ represent the total resistance, inductance, and capacitance, respectively. 
Then the resonant frequency $f_0$ can be expressed as $1/2\pi\sqrt{L_{\text{total}}C_{\text{total}}}$ \cite{23}.
The Q-factor can be modeled based on the equivalent circuit model, as described in \cite{22}:
\begin{equation}
    \begin{aligned}
        Q=\frac{1}{R_{\text{total}}}\sqrt{\frac{L_{\text{total}}}{C_\text{{total}}}} \text{.}
    \end{aligned}
\end{equation} 

The absorption peak depth is determined by the minimum amplitude of scattering coefficient $\Gamma_{\text{min}}$, which 
can be calculated as:
\begin{equation}
    \begin{aligned}
        \Gamma_{\text{min}}=\left|\frac{Z_0-R_{\text{total}}}{Z_0+R_{\text{total}}}\right| \text{.} \label{Gamma_min}
    \end{aligned}
\end{equation}

\begin{figure}[!t]
    \centering
    \includegraphics[width=3.0in]{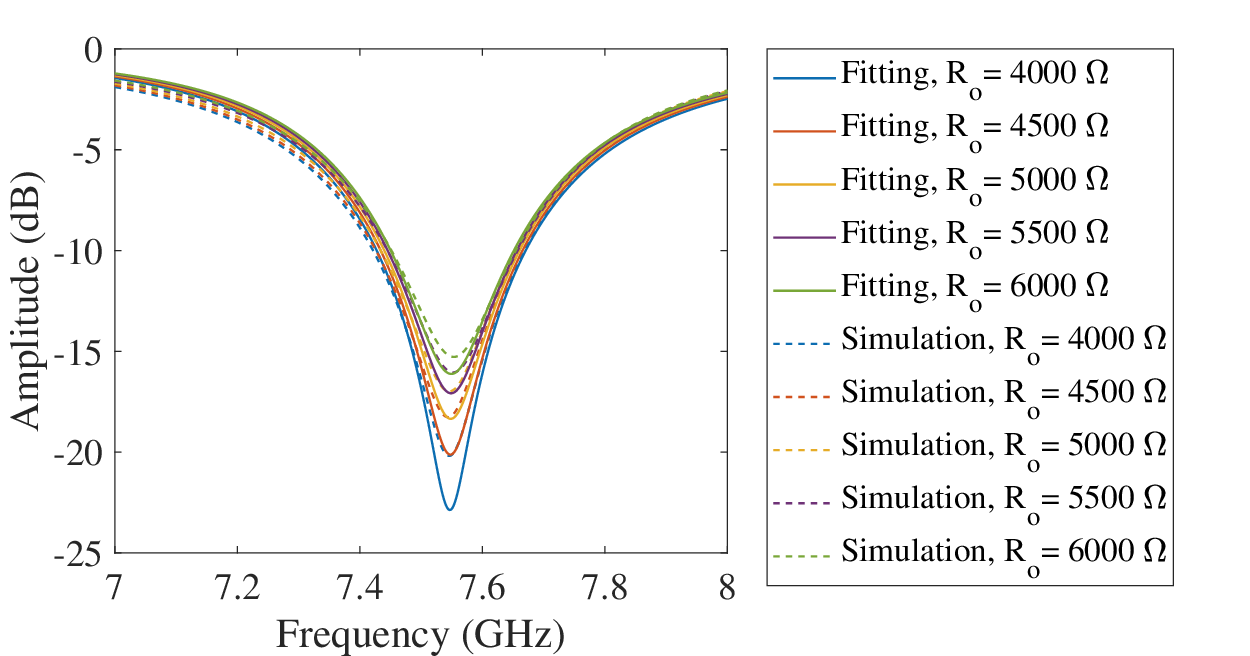}
    \caption{Comparison between simulated results and fitting results with different sensitive material resistance $R_o$.}
    \label{S_para_r}
\end{figure}

Fig. \ref{S_para_r} compares simulated results with equivalent circuit model fitting results for varying sensitive 
material resistance $R_o$. 
As $R_o$ increases, $R_{\text{total}}$ increases correspondingly, resulting in a lower Q-factor. 
Additionally, equation \eqref{Gamma_min} indicates that $\Gamma_{\text{min}}$ is minimized when $R_\text{{total}}=Z_0$. 
Consequently, as $R_o$ increases, $\Gamma_{\text{min}}$ initially decreases and subsequently increases.

\begin{figure}[!t]
    \centering
    \includegraphics[width=3.0in]{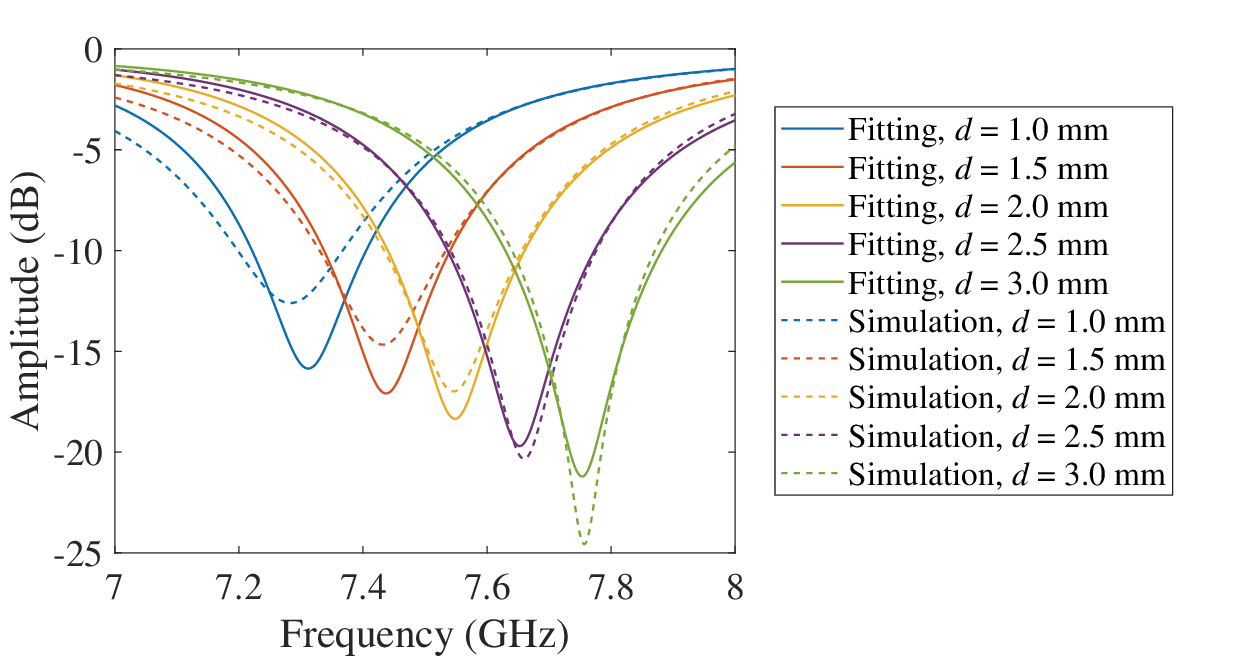}
    \caption{Comparison between simulated results and fitting results with different gap width $d$.}
    \label{S_para_d}
\end{figure}

Fig. \ref{S_para_d} presents a comparison between simulated results and equivalent circuit model fitting results for 
different gap widths $d$. 
An increase in $d$ leads to a decrease in $C_{\text{total}}$ \cite{16}, consequently reducing $C_{\text{total}}$. 
This results in a higher resonant frequency and an increased Q-factor.

\begin{figure}[!t]
    \centering
    \includegraphics[width=3.0in]{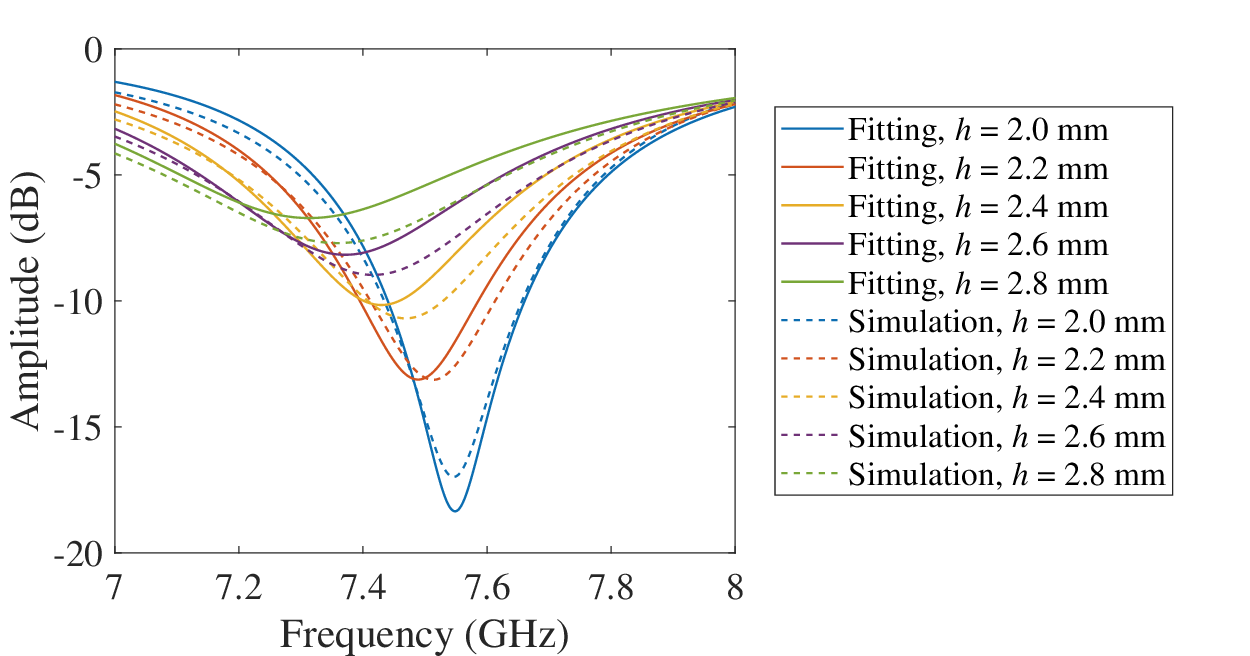}
    \caption{Comparison between simulated results and fitting results with different substrate thickness $h$.}
    \label{S_para_h}
\end{figure}

Fig. \ref{S_para_h} illustrates the comparison between simulated results and equivalent circuit model fitting results 
for varying substrate thickness $h$. 
The substrate impedance $Z_S$ is a complex value with inductive and resistive components \cite{14}. 
As $h$ increases, $Z_S$ increases accordingly, leading to an increase in $L_{\text{total}}$ and a consequent decrease 
in resonant frequency. 
Furthermore, the increase in $h$ results in a higher $R_{\text{total}}$, leading to a lower Q-factor.

\begin{figure}[!t]
    \centering
    \includegraphics[width=3.0in]{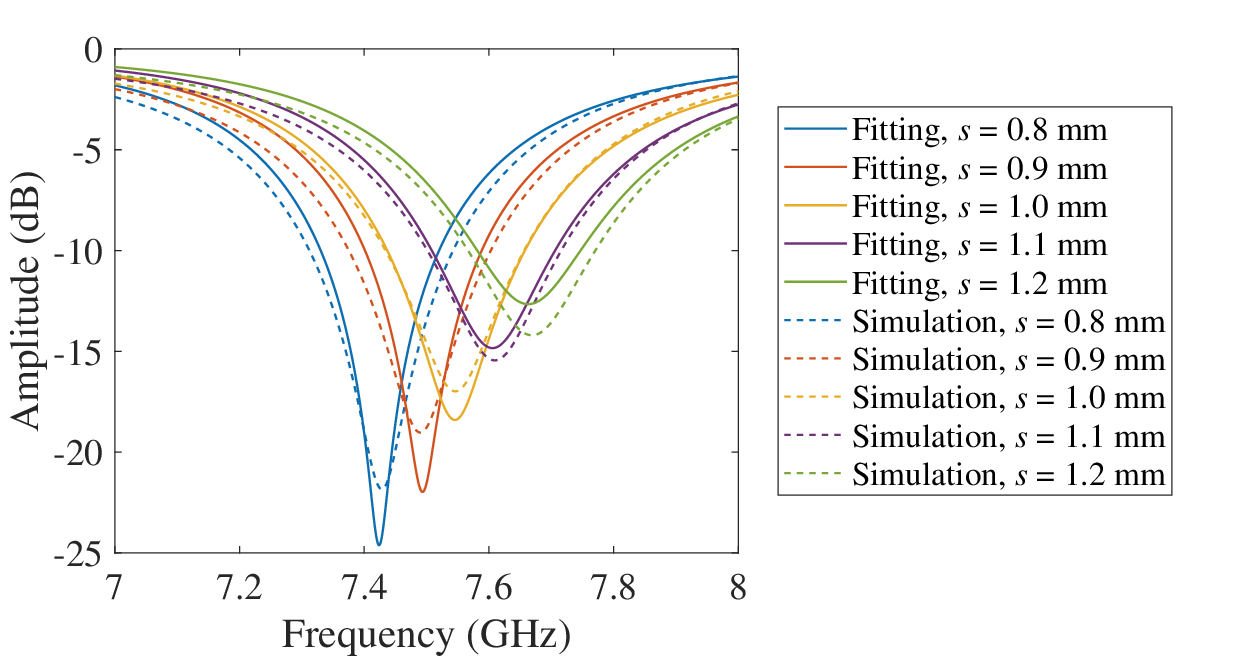}
    \caption{Comparison between simulated results and fitting results with different ring width $s$.}
    \label{S_para_s}
\end{figure}

Fig. \ref{S_para_s} shows the comparison between simulated results and equivalent circuit model fitting results for 
different ring widths $s$. 
An increase in $s$ leads to a decrease in $L_r$, as the average value of the inner and outer ring diameters decreases
\cite{17}. 
This results in a decrease in $L_{\text{total}}$, leading to a higher resonant frequency and a lower Q-factor.

Based on the above discussion, The performance of an SRR is governed by the following design rules:
\begin{itemize}
\item \textbf{Operating Frequency:} Increased with a larger gap width $d$, a thinner substrate $h$, or a wider ring 
width $s$.
\item \textbf{Q-factor:} Enhanced by a larger gap width $d$, a thinner substrate $h$, or a narrower ring width $s$.
\item \textbf{Absorption Peak Depth:} Primarily determined by selecting a sensitive material with an optimal sheet 
resistance $R_o$.  
\end{itemize}

\subsection{Other Tag Designs}
\begin{table*}[!t]
    \belowrulesep=0pt
    \aboverulesep=0pt 
    \renewcommand{\arraystretch}{1.4}
    \doublerulesep 2.2pt
    \caption{List of Other Related Metamaterial Tag Design}
    \label{table1}
    \footnotesize
    \begin{tabular}{|c|p{0.2\textwidth}|p{0.6\textwidth}|}
        \hline
        \makebox[0.1\textwidth][c]{\textbf{Reference}}&\makebox[0.2\textwidth][c]{\textbf{Tag Feature}}
        &\makebox[0.6\textwidth][c]{\textbf{Design Objective}}\\
        \hline
        Ref. [59]&Edge-coupled SRR&Coupling capacitance between the inner and outer ring in edge-coupled SRRs 
        leads to lower resonance frequency, making them more suitable for low frequency applications\\
        \hline
        Ref. [60]&Nested SRR&Numerous gaps in the nested structure significantly increase equivalent 
        capacitance, thereby reducing resonance frequency for low frequency application\\
        \hline
        Ref. [61]&Broadside-coupled SRR&Broadside-coupled SRRs exhibit superior isotropy compared to 
        standard SRRs for normally incident electromagnetic waves, making them more suitable for polarization-independent 
        applications\\
        \hline
        Ref. [62]&Dipole&A series of dipoles in different orientations provides sensitivity for 
        angular rotation monitoring\\
        \hline
        Ref. [63]&SRR with a cantilever at the gap&A sensing medium with environmentally sensitive 
        permittivity is incorporated between two common substrate layers to enhance sensing capabilities\\
        \hline
        Ref. [66]&Multi-layer substrate&Incorporating a sensing medium with environmentally sensitive 
        permittivity between two common substrate layers to enhance sensing capabilities\\
        \hline
        Ref. [67]&Flexible substrate of Kirigami structure&The Kirigami structure transforms 
        one-dimensional strains into three-dimensional rotations, thereby improving strain sensitivity\\
        \hline 
        Ref. [69]&Varactor diode placed on the gap of SRR&Changing the bias voltage changes the impedance 
        in the equivalent circuit, resulting in changes in structure's operating band\\
        \hline 
        Ref. [70]&SRR with adjustable inner ring-substrate distance&By adjusting inner ring-substrate 
        distance using MEMS, structure's operating band is tunable over a wide frequency range\\
        \hline 
    \end{tabular}
\end{table*}
Various studies have proposed alternative tag designs to achieve specific application objectives beyond the most common 
configurations. 
Several researchers have focused on modifying the metal structure's shape to attain desired electromagnetic properties. 
For instance, researchers found that the coupling capacitance between inner and outer ring in edge-coupled SRRs leads 
to lower resonance frequency, making them more suitable for low frequency applications \cite{15}. 
Similarly, a nested SRR design was introduced where numerous gaps in the structure significantly increase equivalent 
capacitance, thereby reducing resonance frequency for low frequency applications \cite{25}. 
It has also been demonstrated that broadside-coupled SRRs are less sensitive to the polarization of incident 
electromagnetic waves compared to standard SRRs, making them more convenient for deployment \cite{28}. 
Other researchers have also specially designed the metal structure to sense specific targets. 
For example, a series of dipoles in different orientations was utilized for angular rotation monitoring \cite{32}, 
while SRR with a cantilever at the gap was adopted for strain sensing \cite{33}.

Another approach is to focused on designing the substrate. 
Some researchers have chosen specialized substrate materials to sense specific targets. 
For instance, ceramic \cite{35} for temperature sensing, polyvinyl alcol \cite{34} and hydrogel \cite{26} for humidity 
sensing. 
Some researchers focus on structures of the substrate. 
Researchers employed a multi-layer substrate, incorporating a sensing medium with environmentally sensitive permittivity 
between two common substrate layers to enhance sensing capabilities \cite{29}. 
Furthermore, a Kirigami structure was designed on a flexible substrate, leveraging its ability to transform 
one-dimensional strains into three-dimensional rotations, thereby improving strain sensitivity \cite{27}.

Researchers have also incorporated tunable components into tag structure to achieve tunability \cite{36}, particularly 
for multi-band operations. 
One approach involves strategically placing tunable electronic elements within the tag. 
For instance, varactor diodes were mounted on the gap of SRR \cite{30}.
By altering the bias voltage of these varactor diodes, the impedance in the equivalent circuit changes, resulting in 
shifts in the tag's operating band. 
Another approach involves mechanically adjusting the position or orientation of tag's fractions. 
An edge-coupled SRR was designed with an adjustable inner ring-substrate distance \cite{31}.
By modifying this distance using MEMS, the structure's operating band can be tuned across a wide frequency range.

\section{Signal Transmission and Detection}
\label{Signal Transmission and Detection}
In the previous sections, we discussed the basic concepts of backscatter tags and their design principles. 
However, sensing and transmission occur simultaneously in this system, creating interdependent challenges. 
First, the sensing tag design affects both sensing and transmission capabilities through its frequency response 
characteristics. 
Second, transmission effects must be considered during the sensing process. 
Consequently, extracting sensing information from wireless signals remains an open research problem. 
In this section, we first introduce a signal sensing and transmission model for metamaterial tags. 
Based on this model, we then discuss key techniques for signal detection.

\subsection{Signal and Interference Model}
The backscatter system is composed of two components, i.e., multiple metamaterial tags and a multi-antenna 
wireless transceiver i.e. Tx and Rx.
In such a system, the Tx and Rx have line-of-sight (LOS) paths to the $i$-th metamaterial tag and can be 
analysis by the Friis' free-space link model~\cite{37}:
\begin{equation}
    \begin{aligned}
        P_{\text{rec},i} (f, \bm{\psi}_{\text{env}})=\frac{P_{\text{Tx}}\sigma \lambda^2}{32\pi^3 r_{\text{Tx},i}^2
        r_{\text{Rx},i}^2}\cdot \Gamma^2(f,\bm{\psi}_{\text{env},i}) \cdot G_{\text{Tx},i}\cdot G_{\text{Rx},i} \text{,}
    \end{aligned}
\end{equation}
where $P_{Tx}$ denotes transmit power, $\sigma$ denotes the area of a metamaterial tag, $f$ is the frequency of the 
incident signal, $r_{\text{Tx},i}$, $r_{\text{Rx},i}$ are the distance between Tx or Rx antenna array and the $i$-th 
tag, $\Gamma(f, \bm{\psi}_{\text{env},i})$ indicates the scattering coefficient of the $i$-th tag defined in 
\ref{scattering coefficient}, $\lambda$ is the wavelength of the incident signal, and $G_{\text{Tx},i}$, 
$G_{\text{Rx},i}$ is the gain factor of the Tx or Rx antenna array in direction of the $i$-th tag, respectively.

In the system, the transmitted signals are not only reflected by the target tag but also by other tags and 
environmental scatters.
Considering the total number of tags to be $I$, the interference received during measurement of the $i$-th metamaterial 
sensing tag can be be expressed as:
\begin{equation}
    \begin{aligned}
        P_{\text{inf},i} (f, \bm{\psi}_{\text{env}}) = &\sum_{j\in [1,I]}^{j\ne i} \frac{P_{\text{Tx}}\sigma \lambda^2}
        {32\pi^3 r_{\text{Tx},j}^2 r_{\text{Rx},j}^2}\cdot \Gamma^2(f, \bm{\psi}_{\text{env},j}) \\ & 
        \cdot G_{\text{Tx},j}\cdot G_{\text{Rx},j} + \eta\cdot P_{\text{Tx}}\cdot \Gamma_{\text{env},i} \text{,}	
    \end{aligned}
\end{equation}
where the first part in the expression is due to the signals reflected by other tags while the second part is due to 
ambient scattering. 

\subsection{Signal Detection Methods}
From the transceiver perspective, the special simultaneous transmission and sensing of a meta-backscatter system requires 
specialized signal processing methods. 
The sensing principle of metamaterial tags relies on frequency response variations of individual tags. 
However, during transmission, the signal is affected not only by the tag's frequency response but also by variable and 
complex wireless channel conditions, as described in signal and interference model. 
Unlike conventional tags that separate sensing and transmission functions using linear and explicit transfer functions, 
meta-backscatter system requires transfer functions that account for wireless channel effects, 
necessitating sophisticated signal processing approaches.
There are two main categories of solutions: traditional signal processing methods and deep learning methods.

\subsubsection{Traditional signal detection methods}
Traditional signal detection methods can be classified as either calibrated or uncalibrated, based on their requirement 
for background normalization---a process involving the subtraction of empty measurements from tag measurements.
Calibrated methods rely on the Singularity Expansion Method (SEM) \cite{o2}, which is premised on the fact that an 
object's electromagnetic response is fully characterized by singularities in the complex frequency plane. 
These singularities are independent of various transmission parameters but depend solely on the scattering tag's 
properties \cite{b8}. 
Thus, sensing information can be separated from the transmitted signal by extracting these singularities. 
Specifically, the Matrix Pencil Method (MPM) \cite{62} and its variant, the Short-Time Matrix Pencil Method (STMPM) 
\cite{63}, were developed for this purpose.

However, in practical scenarios involving rapidly varying background clutter, accurate background normalization becomes 
considerably difficult to achieve. 
Instead of relying on background subtraction, uncalibrated methods focus on disentangling the overlapped components 
within the composite backscattered signal in the time, frequency, or time-frequency domain. 
For instance, a time-gating algorithm was introduced to detect a tag based on a co-polar patch resonator integrated 
with delay stubs \cite{64}. 
Similarly, a joint time-domain and frequency-domain (TD-FD) analysis was presented to effectively separate the 
individual contributions within the backscattered signal \cite{65}.
Expanding on time-frequency analysis, a method utilizing the short-time Fourier transform (STFT) was proposed for 
improved tag detection \cite{66}.
Furthermore, a mathematical model was developed to precisely extract the tag's information from the raw signal \cite{67}.
These methods collectively demonstrate the potential of signal processing techniques in directly isolating tag responses 
under challenging conditions.

\subsubsection{Deep learning methods}
Deep learning methods offer another promising solution due to their capacity to model complex relationships with large 
datasets \cite{51}. 
Specifically, deep learning approaches extract critical features from signals through data-driven learning and map these 
features to sensing outcomes \cite{50}. 
From a network architecture perspective, these models capture wireless channel effects through extensive received signal 
analysis, enabling sensing results to be derived from interference-corrupted signals. 

Deep learning methods can be categorized into two approaches: supervised and unsupervised learning. 
Supervised learning requires labeled training data containing signal measurements and their corresponding ground-truth 
sensing results to train the network as a transfer function, thereby achieving precise quantitative sensing outputs from 
input signals. 
For instance, Convolutional Neural Networks (CNNs) \cite{72} are based on local connectivity and weight-sharing, while 
Recurrent Neural Networks (RNNs) \cite{73} and Long Short-Term Memory (LSTM) networks \cite{b9} model temporal 
dependencies for sequential signal processing. 
Conversely, unsupervised learning identifies patterns and distinctions within unlabeled measurement signals without 
explicitly optimizing for precise results through predefined transfer functions.
Unsupervised learning employs various approaches, ranging from Autoencoders \cite{c4} that discover structures by 
reconstructing inputs to generative models like RBMs \cite{68}, DBNs \cite{69}, and GANs \cite{70}, which learn data 
distributions or create new samples.

Deep learning methods offer a distinct advantage over traditional approaches in highly complex and dynamic wireless 
environments where explicit channel modeling is challenging due to their ability to disentangle tag responses from 
severe interference \cite{74}. 
This advantage comes with two main challenges: substantial data requirements and high computational complexity. 
Deep learning methods necessitates large, diverse, and accurately datasets for training, which requires in-situ data 
collection and labeling after deployment.
Moreover, the real-time processing of signals from a massive number of tags places a substantial computational burden 
on the transceiver, where the deep learning algorithms are typically deployed.
This scalability challenge necessitates research into cloud-edge collaborative network architectures to achieve 
efficient allocation of computational tasks between cloud servers and transceivers.

Several studies have applied deep learning techniques to metamaterial sensing signal detection.
An unsupervised deep learning algorithm was developed that processes received signals and outputs anomaly detection and 
localization indicators to effectively mitigate impairments from wireless transmission \cite{24}.
An end-to-end signal processing framework was proposed utilizing deep learning algorithms to extract key signal 
features, such as resonance frequency and Q-factor, for sensing result determination \cite{10}.
Deep learning was employed for joint channel estimation and signal demodulation to address signal distortion caused by 
variable frequency responses of tags \cite{49}.
These deep learning approaches significantly enhance the signal processing capabilities and information extraction 
performance of meta-backscatter systems.

\section{Implementation and Prototype}
\label{Implementation and Prototype}
In this section, the implementation of a meta-backscatter system prototype are presented, along with experimental results. 
The prototype metamaterial tag was designed following the principles in Section \ref{Metamaterial Tag Design}.
In the prototype transceiver, we use the introduced signal processing method in Section \ref{Signal Transmission and 
Detection} as an example. 

\subsection{Metamaterial Tag Implementation}
The metamaterial unit is composed of the top-metal patterns, the substrate, and the metal ground. 
as shown in Fig. \ref{waveguide}, the top-metal patterns are designed based on the SRR structure whose sizes are 
$10.09\times10.09\times2.4\ \text{mm}^3$, filling the gap with sensitive material.
The gap of of metamaterial units is filled with hygristor TELAiRE HS30P (humidity-sensitive material) to sense 
environmental humidity.
The tag units are designed with specific structure parameters and an absorption peak located around $5.25\ \text{GHz}$.

\subsection{Transceiver platform implementation}
\subsubsection{Transceiver}
The transceiver comprises Tx/Rx antennas and RF devices controlled by a Raspberry Pi.
\begin{itemize}
	\item Tx/Rx Antennas: The Tx/Rx antennas are composed of a waveguide-to-coax adapter (159WCAS) and a rectangular 
    waveguide (159WAL-50) from A-INFO. 
	\item Radio-Frequency (RF) Switches: To reduce the number of RF chains needed in the wireless transceiver, we adopt 
    two RF switches to multiplex single Tx and Rx chains. Specifically, the 8 Tx antennas are connected to a single-
    pole-8-throw RF switch (HMC 321), and the 2 Rx antennas are connected to a single-pole-2-throw RF switch (HMC~270).
	\item Low Noise Amplifiers (LNA): The poles of the two RF switches are connected to the ports of two LNAs 
    (ZX-6043-S+), which can provide an average gain of about $13.5\ \text{dB}$ at $[5.0, 5.5]\ \text{GHz}$ to amplify 
    the transmitted signals.
\end{itemize}

\subsubsection{Signal analyzer}
We employed a USRP hardware platform (NI; X410) to construct the transceiver and provide flexible signal processing 
including signal waveform, filtering, and basic digital analogue conversion.
The transmitter was configured to emit a FMCW signal, while the receiver was set up to down-convert and sampling the 
backscattered signals. 
The FMCW signal is defined by the LabVIEW software in USRP with period of $1e^{-5}$s and bandwidth of $200$MHz in the 
based band.

\subsubsection{Data processor}
A host computer is connected to the Raspberry Pi and signal analyzer through an ethernet switch. 
The software program written in Python is used to control and communicate with them. 
A real time signal pretreatment method runs on the host computer performing the signal filtering, time-frequency 
transformation and basic tag detection algorithm.
Besides, the signal processing algorithm is written in Python with Tensorflow 2.6 framework.

\subsection{Measurement and Results}
The measurement environment is designed to evaluate the performance of the meta-backscatter system in 
terms of the tag's sensing sensitivity.
The experimental setting and results are shown as following:

\subsubsection{Metamaterial tag sensing sensitivity}
\begin{figure}[!t]
    \centering
    \includegraphics[width=3.0in]{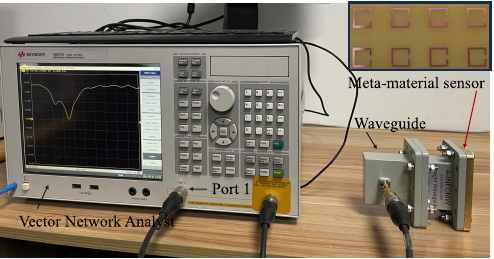}
    \caption{The waveguide measurement scenario \cite{37}.}
    \label{waveguide}
\end{figure}
\begin{figure}[!t]
    \centering
    \subfloat[]{\includegraphics[width=1.7in]{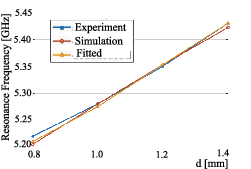}\label{unit1}}
    \subfloat[]{\includegraphics[width=1.7in]{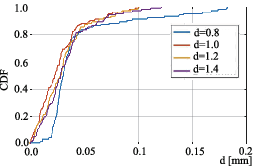}\label{unit2}}
    \caption{(a) Resonance frequencies of fitted, simulated, and experimental scattering coefficients of the 
    metamaterial tags with different gap widths. 
    (b) Cumulative distribution function of the difference between fitted and experimental results \cite{37}.}
    \label{unit_result}
\end{figure}
In this experiment, we fabricated metamaterial tags using the structural parameters defined in Section 
\ref{Metamaterial Tag Design} with varying gap widths and measured their scattering coefficients in a waveguide 
test environment, as shown in Fig. \ref{waveguide}. 
The waveguide environment is a widely adopted method for measuring the frequency response of microwave tags, as it 
confines microwave energy propagation to a single direction within a specified frequency range. 
Scattering coefficient measurements were performed using a square waveguide (WR159) connected to the port of a vector 
network analyzer (VNA) (Agilent E5071C) configured for $S_{11}$ parameter measurement.

Specifically, we characterized metamaterial tags with gap widths of 0.8, 1.0, 1.2, and 1.4 mm, with results presented 
in Fig. \ref{unit_result}. 
As demonstrated in Fig. \ref{unit1}, the resonance frequency increases with increasing gap width $d$, consistent with 
the theoretical predictions described in Section \ref{Metamaterial Tag Design}. 
Furthermore, the maximum deviation in resonance frequency among fitted, simulated, and experimental results is less than 
0.015 GHz. 
The cumulative distribution function shown in Fig. \ref{unit2} reveals an average Euclidean distance of approximately 
0.0438 between fitted and experimental results. 
These results validate the effectiveness of the proposed equivalent circuit model for optimizing metamaterial 
structures.

\subsubsection{Prototype meta-backscatter system}
\begin{figure}[!t]
    \centering
    \includegraphics[width=3.0in]{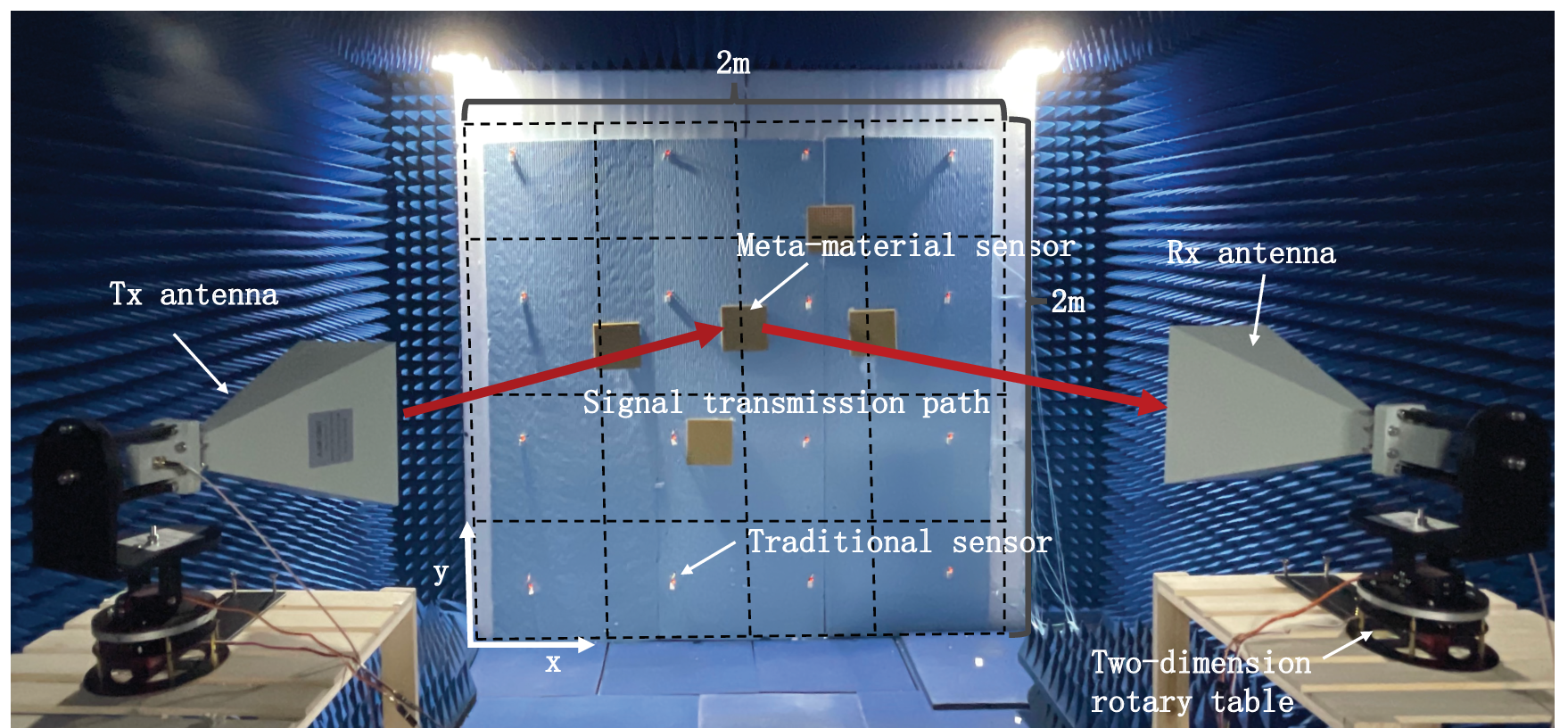}
    \caption{The experimental setting scenario \cite{37}.}
    \label{scenario}
\end{figure}
\begin{figure}[!t]
    \centering
    \subfloat[]{\includegraphics[width=1.6in]{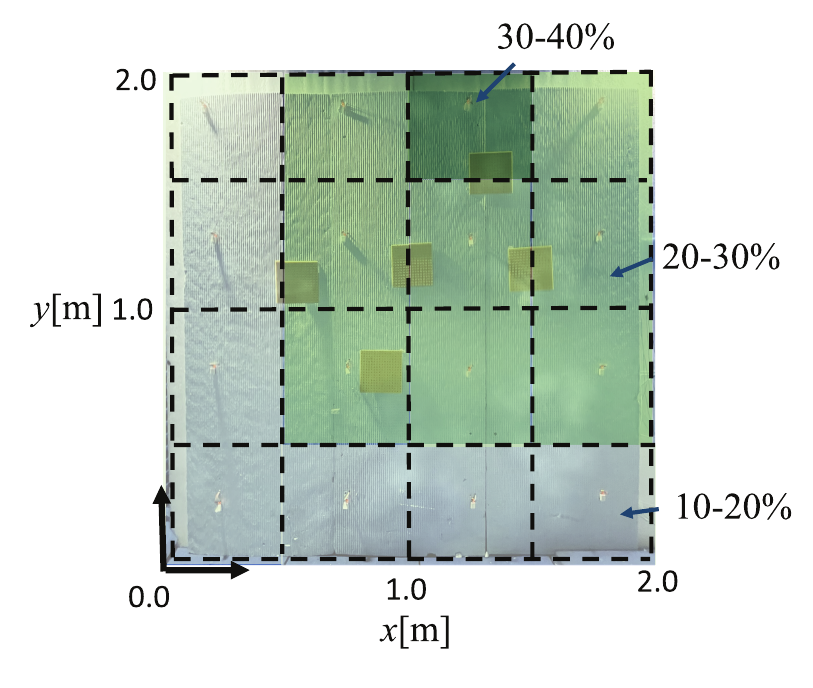}\label{result1}}
    \subfloat[]{\includegraphics[width=1.5in]{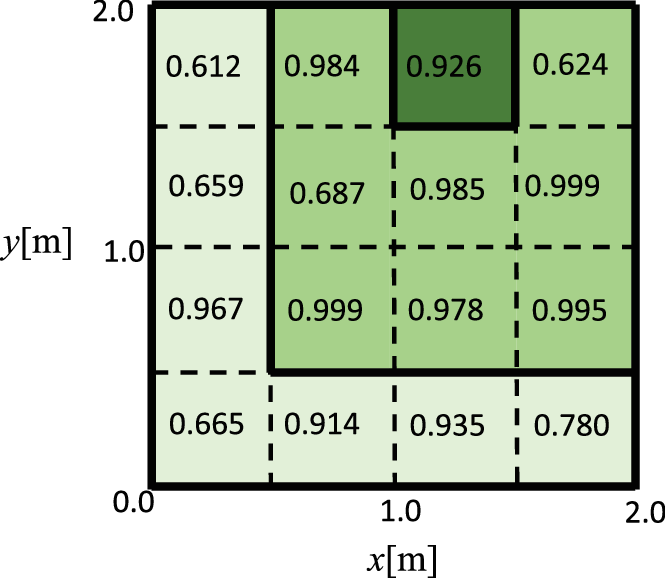}\label{result2}}
    \caption{(a) The humidity category of each space gird. 
    (b) The estimated probability for the correct category \cite{37}.}
    \label{actual_result}
\end{figure}
\begin{figure}[!t]
    \centering
    \subfloat[]{\includegraphics[width=1.6in]{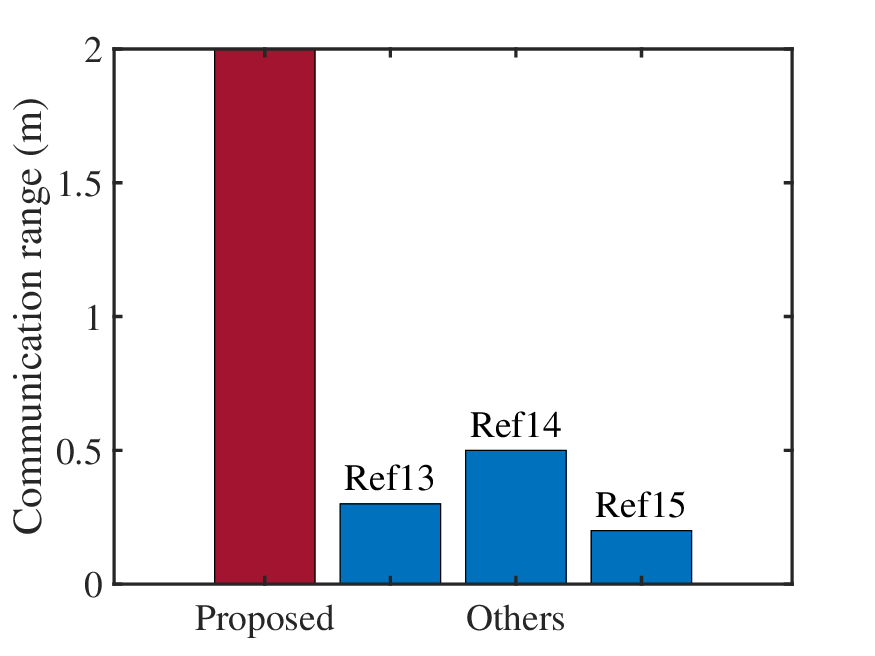}\label{range}}
    \subfloat[]{\includegraphics[width=1.6in]{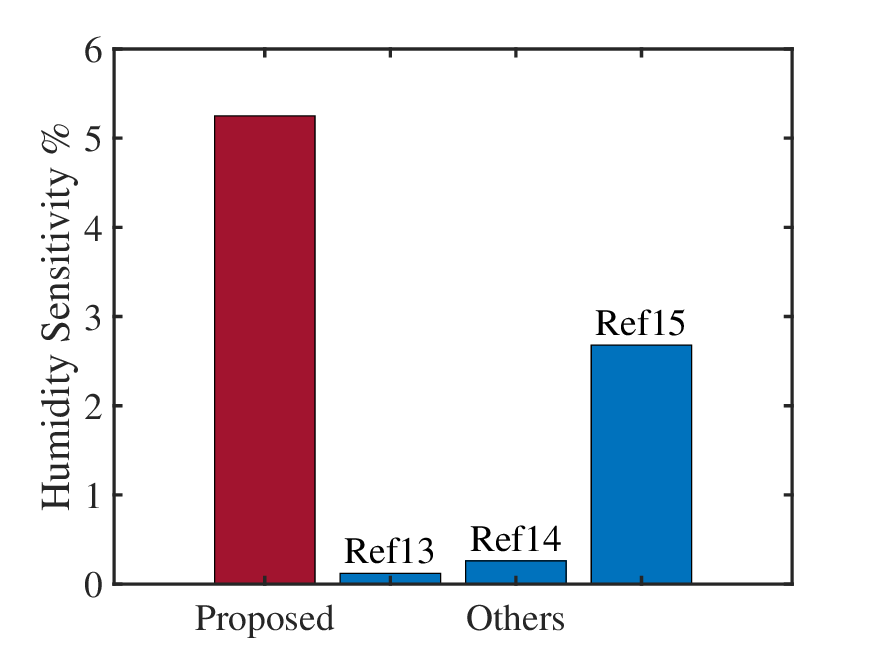}\label{sensitivity}}
    \caption{Comparison between the proposed meta-backscatter system with existing works. (a) Communication range. 
    (b) Sening sensitivity.}
    \label{comparison}
\end{figure}
\begin{table*}[!t]
    \belowrulesep=0pt
    \aboverulesep=0pt 
    \renewcommand{\arraystretch}{1.4}
    \doublerulesep 2.2pt
    \caption{Comparison of the Proposed System with Existing Works}
    \label{table2}
    \footnotesize
    \begin{tabular}{|c|c|c|c|c|c|c|}
        \hline
        \makebox[0.1\textwidth][c]{\textbf{Reference}}&\makebox[0.1\textwidth][c]{\textbf{Frequency Band}}
        &\makebox[0.18\textwidth][c]{\textbf{Test Environment}}&\makebox[0.12\textwidth][c]{\textbf{Sensor Scale}}
        &\makebox[0.1\textwidth][c]{\textbf{Deployment}}&\makebox[0.12\textwidth][c]{\textbf{Transmission Range}}
        &\makebox[0.1\textwidth][c]{\textbf{Sensitivity}}\\
        \hline 
        Proposed&5.0-5.5 GHz&Indoor environment&$10.09\times10.09\ \text{mm}^2$&Sensor array&2.0 m&5.25\%\\
        \hline
        Ref. [13]&6.5-9.0 GHz&Air-tight chamber&$15.0\times6.8\ \text{mm}^2$&Single sensor&0.3 m&0.12\%\\
        \hline
        Ref. [14]&3.0-4.0 GHz&Anechoic chamber&$40.0\times40.0\ \text{mm}^2$&Single sensor&0.5 m&0.26\%\\
        \hline
        Ref. [15]&2.6-3.0 GHz&Waveguide&$30.0\times19.0\ \text{mm}^2$&Single sensor&0.2 m&2.68\%\\
        \hline
    \end{tabular}
\end{table*}
The prototype system was evaluated in a typical indoor environment to demonstrate humidity distribution estimation on a 
wall surface, as illustrated in Fig. \ref{scenario}. 
Specifically, The sensing area covers a $2\ \text{m} \times 2\ \text{m}$ surface discretized into a $4 \times 4$ grids. 
Humidity levels ranging from 0\% to 100\% were categorized into 10 discrete intervals of 10\% each.
To generate controlled humidity distributions, water-retaining materials with varying moisture content were strategically 
positioned across the test surface. 
We adopted a deep learning method based on the preceding bidirectional recurrent network to realize the perception of 
environmental distribution.

Fig. \ref{actual_result} presents representative experimental results for the $4 \times 4$ spatial grids. 
Fig. \ref{result1} illustrates the measured humidity distribution across the grid points, while Fig. \ref{result2} 
displays the classification confidence calculated by the reconstruction algorithm. 
These experimental results demonstrate that the proposed meta-backscatter system exhibits 
high-precision capabilities for humidity distribution detection.
The comparison of the communication range and sening sensitivity with existing backscatter tags is presented in Fig. 
\ref{range} and \ref{sensitivity}. 
A detailed comparison of the measurement setups is listed in Table \ref{table2}.
It is shown that the proposed meta-backscatter system has higher sensitivities and achieves at least a 4-fold 
transmission range compared with the existing works.
Moreover, experiments confirm that the system is capable of detecting humidity anomalies over a maximum distance of 10 
meters, significantly outperforming existing schemes confined to decimeter-level ranges. 

\section{Challenges and Open Questions}
\label{Challenges and Open Questions}
Although metamaterial tags represent a promising approach for backscatter communication, realizing their full 
potential requires addressing several key challenges. 
This section examines three critical aspects: beamforming design in transmitters, multi-tag networking implementation, 
and the utilization of existing communication signals for integrated sensing and communication. 
By analyzing these challenges and discussing potential solutions alongside open research questions, we establish a 
roadmap for advancing metamaterial sensing research and maximizing its transformative impact on BF-IoT systems.

\subsection{Transmitter Beamforming}
Generally, a transmitter equipped with beamforming capability can concentrate signal power toward the target tag to 
maintain adequate signal to-noise ratio (SNR), thereby reducing interference. 
However, the transmitter faces a fundamental constraint: metamaterial tags operate in passive mode and cannot 
provide feedback to assist channel estimation. 
Given this physical limitation, the transmitter must employ an iterative transmission approach to achieve precise 
sensing results. 
The process operates as follows: First, the transmitter sends a signal that is received and processed to generate a 
sensing result, which is then fed back to the transmitter along with the received signal characteristics. 
Second, the transmitter analyzes both the received signal and sensing result to estimate channel conditions and designs 
an optimized signal spectrum for subsequent transmission. 
Finally, using the derived signal spectrum and estimated channel conditions, a broadband beamforming algorithm 
calculates a new beam direction, initiating the next transmission cycle.

\subsection{Multi-tag Networking}
As the demand for IoT sensing continues to grow, deploying multiple tags in a target area to acquire high-resolution 
environmental information becomes increasingly essential. 
However, simultaneous sensing with multiple tags introduces several challenges. 
When multiple tags are active simultaneously, their backscatter signals overlap at the receiver, potentially causing 
significant interference, which may degrade overall system performance. 
Additionally, since meta-backscatter systems operate in the frequency domain, traditional time-division 
multiple access (TDMA)-based protocols for identification are not suitable, making it difficult to distinguish the 
contributions of individual tags to the received signal. 
These issues are particularly problematic in dense deployments, where the number of tags increases significantly. 
To mitigate interference and enable effective networking, it is crucial to develop effective interference management 
strategies for multi-tag networks, thus ensuring accurate sensing and identification of each tag.
To enable effective multi-tag identification, a joint time-frequency domain analysis framework was proposed to mitigate 
interference in multi-tag scenarios \cite{c3}.
A joint time-frequency domain detection algorithm is developed to enhance identification accuracy. 
Additionally, the authors analyzed the error probability of multi-tag identification and found that there exists an 
optimal transmission power that maximizes energy efficiency.

Although notable progress has been made in multi-tag networking, several challenges remain unaddressed. 
First, existing networking protocols lack adaptive access capabilities. Since tag access depends on the reader's 
excitation signals, the management of newly joining tags remains unresolved. 
How to design a dynamic access protocol that can dynamically adjust frequency-domain resource allocation to cope with 
environmental uncertainties remains an open problem.
In addition, existing studies primarily assume static environments. 
In practical applications, however, tag mobility and environmental variations may lead to the dynamic changes in 
interference characteristics, which makes the static models unreliable. 
Developing adaptive interference models that can account for such dynamics remains an open problem.

\subsection{Integrated Sensing and Communication Design}
As discussed in the previous sections, metamaterial tags encode the sensing information into their frequency 
response, which is then captured by the transceiver via signal reflection.
Simultaneously, by functioning as environmental scatterers, metamaterial tags hold the potential to provide additional 
paths for existing communication systems, thereby enhancing the communication performance \cite{60}. 
Consequently, this dual functionality establishes a win-win Integrated Sensing and Communication (ISAC) paradigm. 
Different from traditional ISAC systems that primarily limited to radar-based sensing of geometric and movement 
characteristics \cite{61}, the metamaterial tag-enabled system supports diverse sensing functionalities for ISAC.  
In addition, the ISAC paradigm significantly reduces the extensive bandwidth and hardware requirements of metamaterial 
tags through the efficient reuse of existing communication system resources.

However, the integration of sensing and communication introduces a fundamental trade-off between sensing accuracy and 
communication performance, which are inherently constrained by limited resources, including hardware capabilities, 
available spectrum, and energy budgets. 
To effectively address this challenge and realize the full potential of metamaterial tag-enabled ISAC systems, a 
comprehensive joint optimization framework is essential. 
This framework must encompass three critical design domains: the structural configuration of metamaterial tags, 
transmitter waveforming and beamforming strategies, and receiver signal processing algorithms, all of which must be 
jointly designed to balance the competing demands of sensing and communication objectives \cite{49}.

\subsection{Manufacturability and Deployability}
Printed Circuit Board (PCB) technology presents an ideal solution for the large-scale manufacture of metamaterial tags, 
as their structure essentially comprises periodic metallic patterns printed on dielectric substrates. 
This compatibility, coupled with low material requirements, allows the use of low-cost standard materials, such as 
copper-clad FR-4 substrates.
Additionally, the environmentally sensitive structure can be implemented using thermistors, hygrometers, or 
photoresistors combined with Surface Mount Technology (SMT).
Leveraging fundamental PCB processes, the unit cost of individual metamaterial tags can be reduced to just a few cents. 
Furthermore, their flat, passive design makes them exceptionally suitable for integration into building structures like 
walls and floor slabs. 
Once embedded during construction, these tags can function as building structural monitors throughout the building 
phase and continue to provide long-term environmental monitoring thereafter. 
Moreover, once deployed within a fixed operational environment, the tags can undergo a one-time calibration. 
This initial sampling and reference setup ensure consistent and reliable long-term measurements without the need for 
recurrent adjustments.

\section{Conclusion}
\label{Conclusion}
This survey has systematically articulated the unique position and design considerations of metamaterial tags within 
the BF-IoT landscape. 
Unlike existing BF-IoT devices, metamaterial tags offer significantly extended communication range, yet they 
introduce the fundamental challenge of jointly designing coupled sensing and transmission functions. 
To address this core complexity, this work has consolidated the underlying principles, design methodologies, and 
implementation techniques into a unified framework, providing a much-needed synthesis of this fragmented field. 
Our prototype experiments demonstrated the effectiveness of meta-backscatter systems for accuracy environmental 
monitoring while significantly extending the communication range. 
Finally, we discussed the key challenges and open research directions, including beamforming, networking, and ISAC design. 
The potential of metamaterial tags to unlock new frontiers in BF-IoT is evident. 
As research advances, we expect them to mature into a key enabling technology for building pervasive and intelligent 
sensing networks.

Conflict of interest statement. None declared.

\bibliography{reference}

\end{document}